\documentclass[sigconf]{acmart}

\setcopyright{none}
\renewcommand\footnotetextcopyrightpermission[1]{}

\usepackage{booktabs}
\usepackage{array}
\usepackage{tabularx}
\usepackage{amsmath}
\usepackage{enumitem}
\usepackage{microtype}
\usepackage{balance}
\usepackage{placeins}
\graphicspath{{figures/}}
\begin{document}

\title{When Do Institutions Beat Intelligence?}

\author{Zhengye Han}
\affiliation{%
  \institution{New York University}
  \department{Department of Electrical and Computer Engineering}
  \city{Brooklyn}
  \state{New York}
  \country{United States}
}
\email{zh3286@nyu.edu}

\begin{abstract}
More capable agents do not necessarily form a more capable collective. A multi-agent system may jointly possess sufficient information yet fail because evidence is poorly routed, unreliable reports enter public belief, correlated claims masquerade as independent support, shared state becomes stale or strategically distorted, or useful evidence is exposed through an ineffective action interface. We ask when additional resources should improve the reasoner and when they should instead change the institutional structure through which the collective forms and acts on public information. Drawing on functional distinctions from research on group decision making and distributed cognition, we construct controlled artificial ecologies around four loci of collective failure: access and routing, admission and dependence, state maintenance and incentives, and representation and action. Across these ecologies, we separately vary model capability and institutional structure, pairing positive interventions with matched reasoning baselines and mechanism-breaking controls. The experiments reveal a consistent boundary: institutions help when they repair failures in how a collective constructs usable public state, but lose their advantage when their signals are uninformative or uncheckable, when stronger intelligence can perform the same transformation directly, or when the resulting state cannot support reliable action. Our results recast the choice between intelligence and institutions as a diagnosis of where collective reasoning fails.
\end{abstract}


\keywords{multi-agent systems, institutions, collective intelligence, group decision making, large language model agents, empirical methodology}

\maketitle

\pagestyle{plain}

\section{Introduction}

More capable agents do not necessarily form a more capable collective.
A system may jointly contain enough information to solve a task and still fail because that information never becomes jointly accessible, because public belief admits unsupported or dependent evidence, because shared state is not maintained, or because the resulting representation cannot support reliable action.
Such failures are not reducible to weakness inside any one agent.
They arise along the path by which private observations become public reasons for collective action.

This distinction creates a concrete design choice.
An \emph{intelligence intervention} strengthens the reasoner, for example by changing model family, scale, or inference budget.
An \emph{institutional intervention} changes the structure through which agents observe, report, verify, update, and act on information.
Intelligence expands what a reasoner can compute from the state it receives.
Institutions change how the collective constructs that state.
Our question is therefore not whether institutions are generally better than intelligence, but \emph{which source of collective failure each intervention can actually repair}.

\begin{figure*}[!t]
  \centering
  \includegraphics[width=0.96\textwidth]{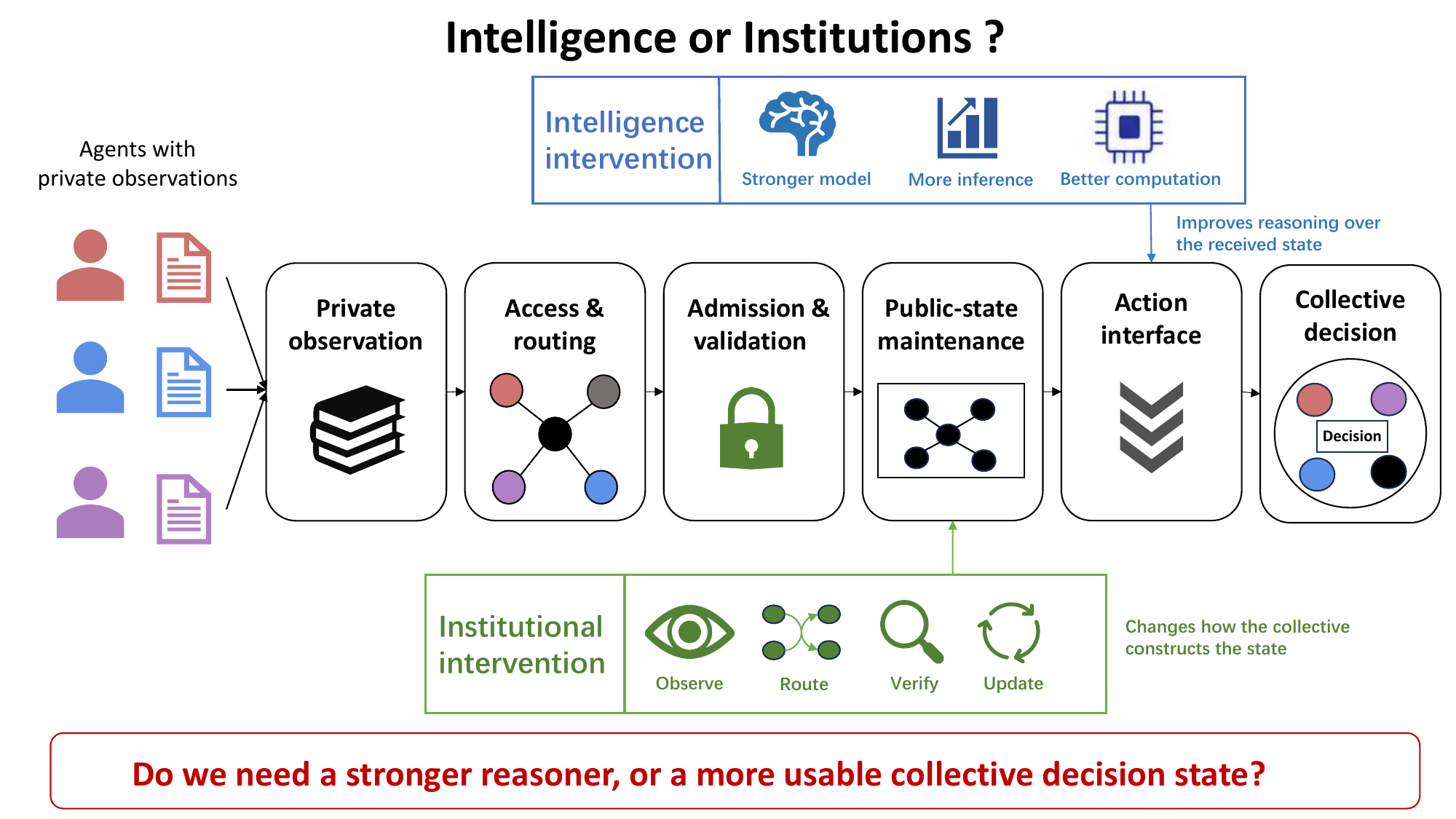}
  \caption{\textbf{Intelligence and institutions intervene at different levels of collective computation.}
  Intelligence interventions strengthen the reasoning applied to a received state.
  Institutional interventions change how private observations become decision-relevant public state through access, routing, admission, verification, updating, and the action interface.
  We ask whether a failing collective requires a stronger reasoner or a more usable decision state.}
  \Description{A left-to-right pipeline maps agents' private observations through access and routing, admission and validation, public-state maintenance, and an action interface to a collective decision. A blue band represents intelligence interventions that improve reasoning over the received state, while a green band represents institutional interventions that change how the collective constructs that state.}
  \label{fig:overview}
\end{figure*}

The ingredients of this distinction are well established in multi-agent systems:
protocols, organizations, norms, monitoring, and shared state regulate interaction among autonomous agents \cite{valckenaers2007}, while institutional analysis emphasizes how rules-in-use shape collective outcomes \cite{ostrom1990,ostrom2009}.
What is harder to identify empirically is \emph{which intervention repaired a failing artificial collective}.
Contemporary agent workflows often vary roles, debate, voting, retrieval, memory, model capability, and call count together \cite{chan2024,du2024,wu2024}.
An improvement in end-task performance then shows that the pipeline changed, but does not reveal whether the binding constraint was reasoning capacity, evidence access, evidential validity, enforceability, or the action interface.
This ambiguity matters in both directions:
institutional machinery introduces additional computation and failure modes, while additional reasoning is wasted when every reasoner continues to receive the same unusable state.

We study this identification problem through controlled artificial ecologies.
Rather than treating environments as interchangeable benchmarks, we design them around four locations at which collective information processing can fail (Fig.~\ref{fig:overview}).
\emph{Access and routing} asks whether distributed knowledge becomes jointly available.
\emph{Admission and dependence} asks which reports become public evidence and whether repeated reports provide independent support.
\emph{Maintenance and incentives} asks whether public state remains current and whether deviations can be observed and made consequential.
\emph{Representation and action} asks whether a constructed state remains useful relative to stronger capability and whether the final agent can reliably act on it.

These distinctions are motivated by recurrent problems in research on human groups, without assuming psychological equivalence between human and artificial agents.
Hidden-profile and transactive-memory research shows that a group may collectively possess decisive information while failing to retrieve and coordinate it effectively \cite{stasser1985,lu2012,stasser2000}.
Research on social influence shows why repeated agreement need not constitute independent evidence \cite{becker2017,freyrijt2021,almaatouq2022}.
Work on accountability emphasizes that consequences depend on what can be observed and attributed \cite{tetlock1989,tetlockboettger1989,schillemans2022}.
Distributed-cognition research, in turn, treats external representations as part of the cognitive system rather than as neutral containers of already-completed reasoning \cite{hutchins1995,zhangnorman1994,scaife1996}.
We use these literatures to motivate distinct functional failure structures, not to claim that artificial agents reproduce human psychological mechanisms.

Each ecology then isolates one such failure and asks whether changing capability or changing institutional structure repairs it.
Institutional interventions are paired with stronger-model or call-matched raw baselines, together with mechanism-breaking controls under which the proposed institutional explanation should lose value.
Call-matched sampling, voting, and debate test whether an apparent gain is merely additional inference.
Below-threshold routing tests whether coverage rather than workflow complexity matters.
Syntax-only validation separates format from evidential validity.
Zero-checkability conditions separate sanctions from enforceability.
Frontier models and learned retrieval test whether capability can substitute for an explicit institution.
Fixed-evidence interfaces separate better evidence construction from better execution.
The objective is therefore not to rank workflows, but to identify \emph{where collective computation fails and why a particular intervention succeeds or becomes redundant}.

We make three contributions.
First, we introduce a functionally defined taxonomy of artificial collective failures, motivated by group decision research, that separates failures of access, public belief, state maintenance, and action.
Second, we develop matched, mechanism-breaking comparisons that distinguish institutional repair from additional reasoning and specify conditions under which an institutional explanation should fail.
Third, across controlled ecologies, strategic interaction, multiple model families, and natural evidence tasks, we identify both institutional advantage and its boundaries:
institutions help when they repair the construction of usable public state, but lose value when their signals are invalid or uncheckable, when stronger capability performs the same transformation directly, or when the resulting state cannot support reliable action.

\section{Institutions as Rules for Collective State Construction}
\label{sec:formalism}

We separate the task ecology from the rules through which a collective
turns private observations into a decision-relevant public state. An ecology is
\[
\mathcal{E}
=
\left\langle
N,S,\{O_i\}_{i\in N},\{A_i\}_{i\in N},
T,\{R_i\}_{i\in N},H
\right\rangle,
\]
with agents \(N\), latent task state \(S\), private observation channels
\(O_i\), reporting or action spaces \(A_i\), transition dynamics \(T\),
possibly heterogeneous rewards \(R_i\), and horizon \(H\).

Within an ecology, we represent an institution as
\[
I=\langle \rho,V,U,\Phi\rangle.
\]
The component \(\rho\) specifies observation and reporting rights:
which agents may access, produce, or route which information.
Given the resulting reports \(r_t\), \(V\) specifies which reports are
entitled to enter public belief. The admitted evidence
\(\tilde r_t = V(r_t)\) updates a shared public state
\(
p_t = U(p_{t-1},\tilde r_t),
\)
and \(\Phi\) determines how that state is represented and exposed to the
final decision process. A reasoner with capability \(m\) then acts on
\(\Phi(p_t)\).

This definition makes the institutional boundary explicit. We call
\(\rho,V,U,\Phi\) institutional because they are externally specified rules
governing collective information rights, admissibility, state maintenance,
and action affordances, rather than parameters of an individual reasoner.
They may be role- or history-dependent and may condition future interaction
on observable behavior. The definition therefore includes routing,
verification, public memory, audit, and institutionally constrained action
interfaces, while excluding additional sampling or cosmetic reformatting
that leaves the information available to the collective and its action rules
unchanged. This operationalization follows the institutional perspective that
rules and coordination structures regulate interaction among agents
\cite{ostrom1990,ostrom2009,valckenaers2007}.

\paragraph{Institutional and capability interventions.}
Let \(Y(e,I,m)\) denote system performance on paired instance \(e\) under
institution \(I\) and reasoner capability \(m\). A capability intervention
changes \(m\) while holding the collective information structure fixed; an
institutional intervention changes how that information is routed, admitted,
updated, or exposed while holding the reasoner fixed.

For regime \(r\in\{+,-\}\), define
\[
\tau_I^r(m)
=
\mathbb{E}_{e\sim\mathcal{E}}
\left[
Y(e,I^r,m)-Y(e,I_0^r,m)
\right],
\]
where \(I^+\) contains the proposed informative institutional signal and
\(I^-\) preserves the workflow while breaking that signal. We define the
mechanism-validity interaction
\[
\Gamma_I(m)
=
\tau_I^+(m)-\tau_I^-(m).
\]
An institutional gain receives mechanistic credit only when performance
tracks the signal the institution claims to exploit: structure alone is
insufficient if the gain persists when that signal is broken.

To compare structure with additional capability, we use the crossover
\[
C_I(m_s,m_\ell)
=
\mathbb{E}_{e\sim\mathcal{E}}
\left[
Y(e,I^+,m_s)-Y(e,I_0,m_\ell)
\right],
\]
where \(m_s\) is a weaker institutionally supported reasoner and \(m_\ell\)
a stronger raw reasoner evaluated on the same paired instances and, where
applicable, under matched call budgets. A positive crossover is therefore
evidence for institutional advantage only when the corresponding
mechanism-validity test is also positive.

\paragraph{Diagnostic failure sources.}
We do not assume that institutional effects admit an additive structural
decomposition. Instead, the experiments distinguish five diagnostic sources
of success or failure: correction of an active bottleneck (\(B\)),
information lost during state construction (\(L\)), mismatch between the
constructed state and downstream execution (\(X\)), institutional overhead
(\(O\)), and substitution when a sufficiently capable learned component
performs the same transformation directly (\(S\)). Broken or uncheckable
signals test \(B\); construction diagnostics expose \(L\); fixed-evidence
interfaces isolate \(X\); explicit resource accounting measures \(O\); and
stronger raw models or learned constructors probe \(S\). These are diagnostic
labels tied to matched controls, not separately identified causal parameters.

\section{Experimental Design}
\label{sec:design}

\textbf{Design logic.} Every main comparison is paired by seed or question. Controlled structural experiments and semantic validation form the causal core; cross-family and zero-checkability tests provide boundary replication; natural tasks probe transfer, substitution, and execution rather than an aggregate benchmark score.

\noindent
\textbf{Construction conditions.} The locked construction core includes raw inference, call-matched sampling, majority vote, a free-form debate proxy, unvalidated pooling, below-threshold institutions, valid institutions, and diagnostic public-state execution.

\noindent
\textbf{Validity and admission designs.} Static validity studies use a $2\times2$ design: the same finalizer sees raw or institutionally constructed state under a valid or deliberately broken regime. Controlled semantic validation fixes the transcript and changes only whether one unsupported row is admitted.

\noindent
\textbf{Repeated and natural-task designs.} The repeated ecology crosses private-signal capability, audit probability, checkability, and sanction strength under common random numbers. Natural tasks keep selected passage count fixed; learned retrieval tests construction substitution, while ordered and pointer interfaces hold evidence fixed and change only the action representation.

\noindent
\textbf{Model and scale coverage.} The locked hosted core evaluates Llama 3.1 8B and Llama 3.3 70B on 50 held-out seeds per construction ecology. Separate panels test Qwen 235B/Mistral 24B family transfer, validity interactions for Qwen 9B, Qwen 235B, and Mistral 24B, and frontier substitution for Claude Sonnet 4.6, Gemini 2.5 Pro, and DeepSeek V3.2.

\noindent
\textbf{Sample sizes.} The repeated-audit grid uses 100 matched seeds per cell, with 50 for the zero-checkability boundary. HotpotQA uses 20 paired questions per model. MuSiQue retrieval uses 300 questions balanced across 2--4 hops; hosted execution uses 21 or 30 questions crossed with three finalizers. The frozen-history comparison uses 20 simulation seeds and three post-burn-in questions per seed.

\noindent
\textbf{Statistics.} We report paired mean differences and nonparametric 95\% bootstrap intervals. Resampling follows the assignment unit: seed for controlled ecologies, question for ordinary natural panels, and connected source-component cluster when dependence remains. The structural panels retain arm means because absolute failure is informative; other main panels emphasize paired effects.

\section{Distributed Knowledge: Access and Routing}
\label{sec:access}

\paragraph{Group-psychology motivation.}
Hidden-profile research separates information that a group \emph{possesses} from information that enters collective deliberation: when critical facts are split across members so that no individual's private view favors the correct option but the group's pooled view does, discussion still gravitates toward information everyone already shares rather than the decisive unique pieces.
Across 65 studies, groups discussed substantially more shared than unique information and were far less likely to solve hidden profiles than groups given complete information \cite{lu2012}.
Transactive-memory work adds a second distinction: coordination improves when members know who has access to which expertise \cite{stasser2000,vanginkel2009}.
The functional lesson is not that artificial agents share a human bias.
It is that distributed possession is not yet collective availability.

\begin{figure}[htbp]
  \centering
  \includegraphics[width=\columnwidth]{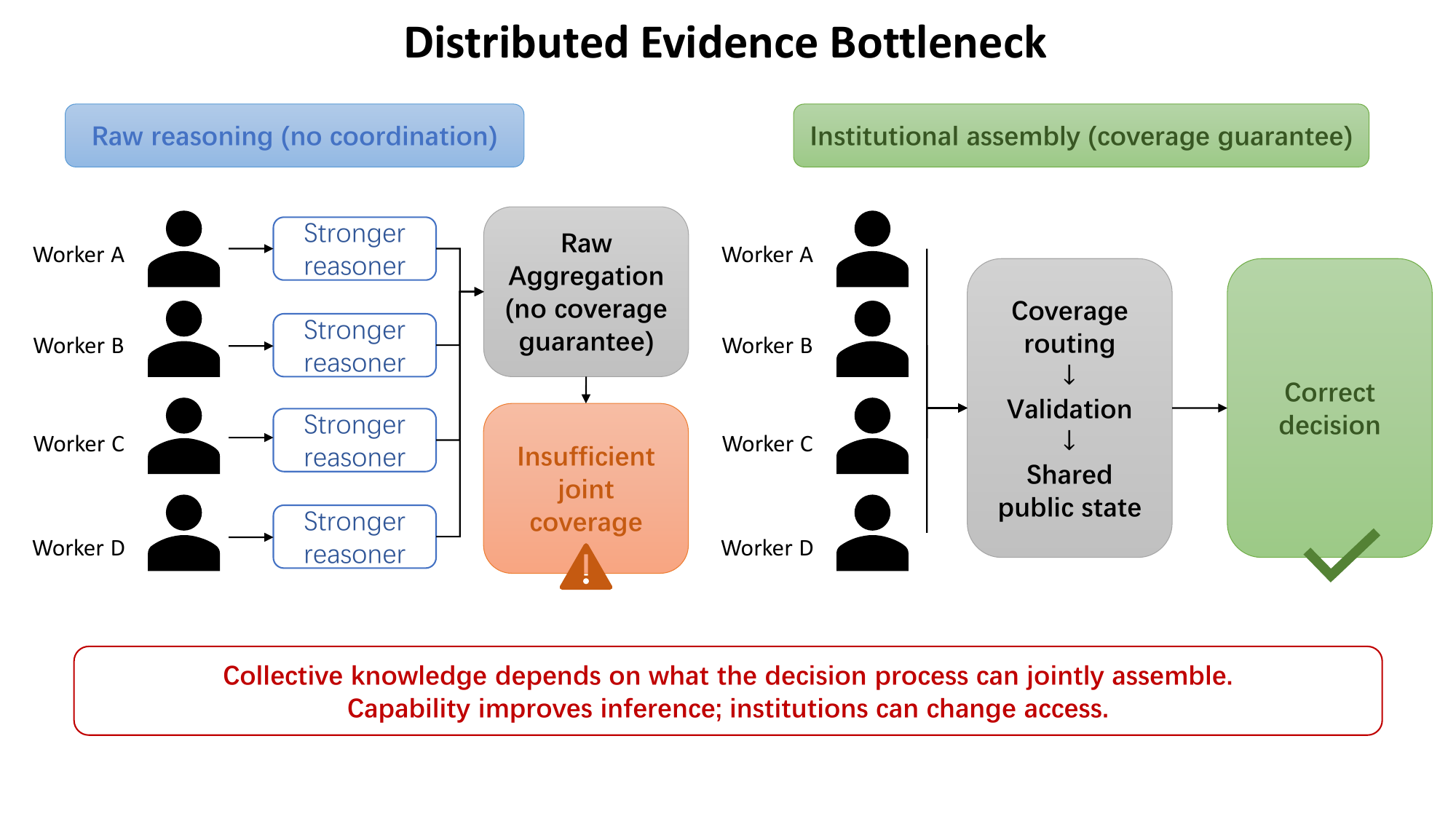}
  \caption{\textbf{Distributed evidence as an access bottleneck.}
  Raw reasoning can spend more capability on partial views without guaranteeing joint coverage.
  Here, \emph{coverage} means that the public decision state contains all complementary evidence components required for the task, rather than merely more reports or more computation.
  Institutional routing and validation instead assemble a jointly sufficient public state by making those complementary components collectively available to the finalizer.}
  \Description{A contrast between uncoordinated stronger reasoning over partial worker views and institutional assembly through coverage routing, validation, a shared public state, and a finalizer. Coverage refers to whether the final public state jointly contains all complementary evidence needed for the decision.}
  \label{fig:distributed}
  \vspace{-0.3cm}
\end{figure}

\paragraph{Artificial ecology.}
Figure~\ref{fig:distributed} turns this distinction into a controlled artificial ecology.
The task has 12 candidate answers (A--L) and a deck of 16 \emph{clue cards}, each encoding one logical constraint on the correct answer -- e.g., a card of type \texttt{eliminated\_set} with options \{K, E\} states that the answer is neither K nor E.
No single card, and no individual worker's local view, is sufficient to determine the answer.
We distinguish three objects: a \emph{clue card} is the private input a worker is shown; a \emph{constraint} is the logical restriction it expresses; and a \emph{record} is the worker's structured transcription of that card (id, type, options).
The institution validates each record against the card the worker actually saw, then aggregates validated records into one public constraint table exposed to the finalizer.
Raw 8B and 70B systems receive the same underlying cards with matched calls.
An assignment $m{\times}k$ gives $m$ workers $k$ cards each.
$3{\times}2$ and $3{\times}4$ use option-symbol-balanced routing that allows cards to repeat across workers and leaves coverage incomplete; only the $4{\times}4$ assignment uses a disjoint partition that covers all 16 cards exactly once and supplies a jointly sufficient state.
A related expertise-routing ecology uses public credentials to direct domain-specific evidence and breaks the credential signal in the control condition.

\begin{figure*}[htbp]
  \centering
  \includegraphics[width=1\textwidth]{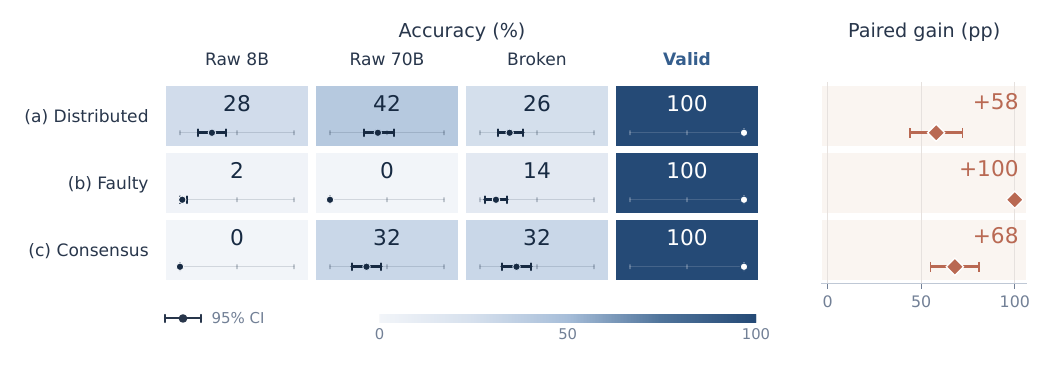}
      \vspace{-0.3cm}
    \caption{\textbf{Institutions overcome bottlenecks that persist under stronger raw reasoning.}
    Rows show distributed evidence, faulty evidence, and wrong consensus.
    Broken and valid institutions use 8B models.
    Cells encode accuracy (\%), with inset intervals on a shared within-cell 0--100 scale.
    Right: paired valid-8B minus raw-70B accuracy in percentage points (pp).
    Whiskers show 95\% confidence intervals.
    Results concern these frozen ecologies.}
  \label{fig:results-structural}
\end{figure*}

\paragraph{Result and interpretation.}
In distributed evidence, the valid five-call 8B institution exceeds the five-call raw 70B system by $+0.58$ [0.44, 0.72] (Fig.~\ref{fig:results-structural}).
As routing becomes more complete, task accuracy rises from $0.26$ under $3{\times}2$, to $0.96$ under $3{\times}4$, and to $1.00$ under the disjoint $4{\times}4$ assignment; correspondingly, mean coverage of decisive clues rises from $0.357$ to $0.760$ to $1.000$, while the validated public table uniquely identifies the answer in $0\%$, $94\%$, and $100\%$ of trials, respectively -- the chain is routing completeness $\rightarrow$ decisive-evidence coverage $\rightarrow$ unique public decision state $\rightarrow$ final accuracy.
Separate Qwen/Mistral probes reproduce the qualitative threshold pattern.
Expertise-routing validity is tested with the cross-model interactions in Fig.~\ref{fig:results-validity}a.
The conclusion is narrower than ``small models beat large models.''
A stronger reasoner can make better use of evidence it receives, but cannot infer constraints that never become jointly available.
The institution helps by changing access to the decision state.
If a capable raw reasoner can directly reconstruct the same transformation, that conclusion should reverse; Section~\ref{sec:representation} tests exactly that boundary.

\begin{figure}[htbp]
  \centering
  \includegraphics[width=\columnwidth]{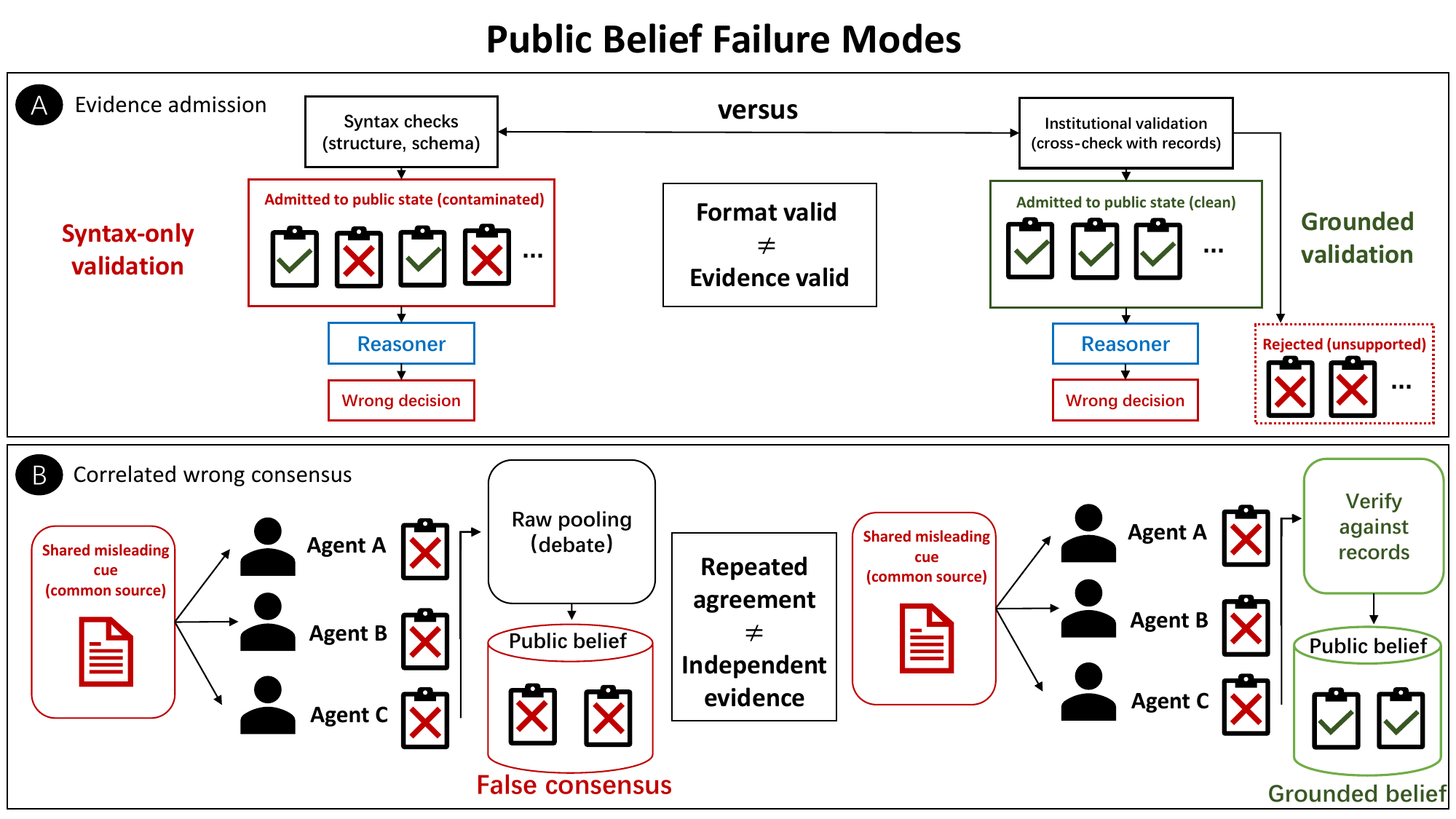}
  \caption{\textbf{Two public-belief failure modes.}
  Schema-valid reports can remain evidentially unsupported, and repeated reports can share one underlying error source.
  Grounded verification changes which reports may update public belief; it does not imply that agreement or voting is generally unreliable.
  Icons denote representative report sets.}
  \Description{Panel A contrasts syntax-only with grounded validation. Panel B contrasts raw pooling of three correlated reports with verification against records that rejects unsupported correlated reports.}
  \label{fig:belief}
  \vspace{-0.3cm}
\end{figure}

\section{Public Belief: Admission and Dependence}
\label{sec:admission}

\paragraph{Group-psychology motivation.}
Communication can fail after evidence has been shared.
Human-group studies show that decision makers may evaluate the same information differently depending on prior preferences \cite{greitemeyer2003}, while work on the wisdom of crowds shows that agreement is informative only under assumptions about the independence and distribution of judgments \cite{becker2017,freyrijt2021,almaatouq2022}.
A repeated claim can therefore increase consensus without adding independent evidence.
The corresponding institutional question is epistemic: which reports are entitled to update public belief?

\paragraph{Artificial ecologies.}
Figure~\ref{fig:belief} separates admission from evidential dependence.
In the faulty-evidence ecology, malformed or logically invalid records enter a common pool.
The institution validates reports against visible record rules before updating public state; raw and call-matched baselines repeatedly reason over the contaminated pool.
The wrong-consensus ecology creates several reports that repeat one salient but incorrect cue.
Voting, repeated sampling, and free-form debate preserve correlation, whereas exact-record verification tests whether claims have independent support.
A controlled semantic-validation experiment isolates admission from generation: one schema-valid report flips a visible constraint type, and the finalizer receives either syntax-only admission or grounded admission based on the exact visible record.

\begin{figure*}[!t]
  \centering
  \includegraphics[width=1\textwidth]{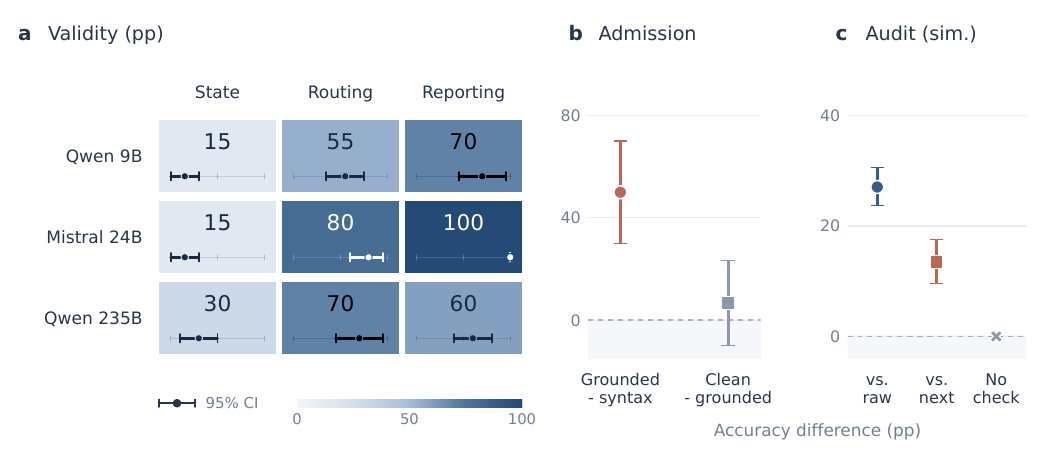}
    \vspace{-0.5cm}
    \caption{\textbf{Institutional effects depend on validity and checkability.}
    (a) Valid-rule gain minus broken-rule gain across public state, expertise routing, and strategic reporting (20 paired seeds per cell).
    (b) Admission contrasts with transcript and finalizer fixed (30 paired seeds).
    (c) Simulated audit gains over unaudited baselines at the same (vs.\ raw) or next capability (vs.\ next); No check denotes zero gain across four sanction levels without observable violations.
    Effects are in pp; whiskers show 95\% confidence intervals, using a shared within-cell 0--100 scale in (a).}
  \label{fig:results-validity}
\end{figure*}

\paragraph{Result and interpretation.}
Validated construction reaches 1.00 in faulty evidence while call-matched raw reasoning remains at 0.00.
Under wrong consensus, the valid 8B institution exceeds the four-call raw 70B baseline by $+0.68$ [0.55, 0.81] (Fig.~\ref{fig:results-structural}).
In the controlled transcript, grounded validation exceeds syntax-only admission by $+0.50$ [0.30, 0.70], while clean minus grounded is only $+0.067$ [$-0.100$, 0.233] (Fig.~\ref{fig:results-validity}b).
Because clean and grounded-filtered finalizer prompts are identical, the gain is rejection of unsupported evidence rather than cleaner prose, extra sampling, or an answer-bearing hint.

These experiments distinguish social agreement from evidential warrant.
Institutions help when they change what is allowed to count as a public reason, not when they merely cause the same evidence to be reconsidered more often.
The broken-rule and syntax-only controls are therefore central: structure without a valid admission signal receives no causal credit.

\begin{figure}[!htbp]
  \centering
  \includegraphics[width=\columnwidth]{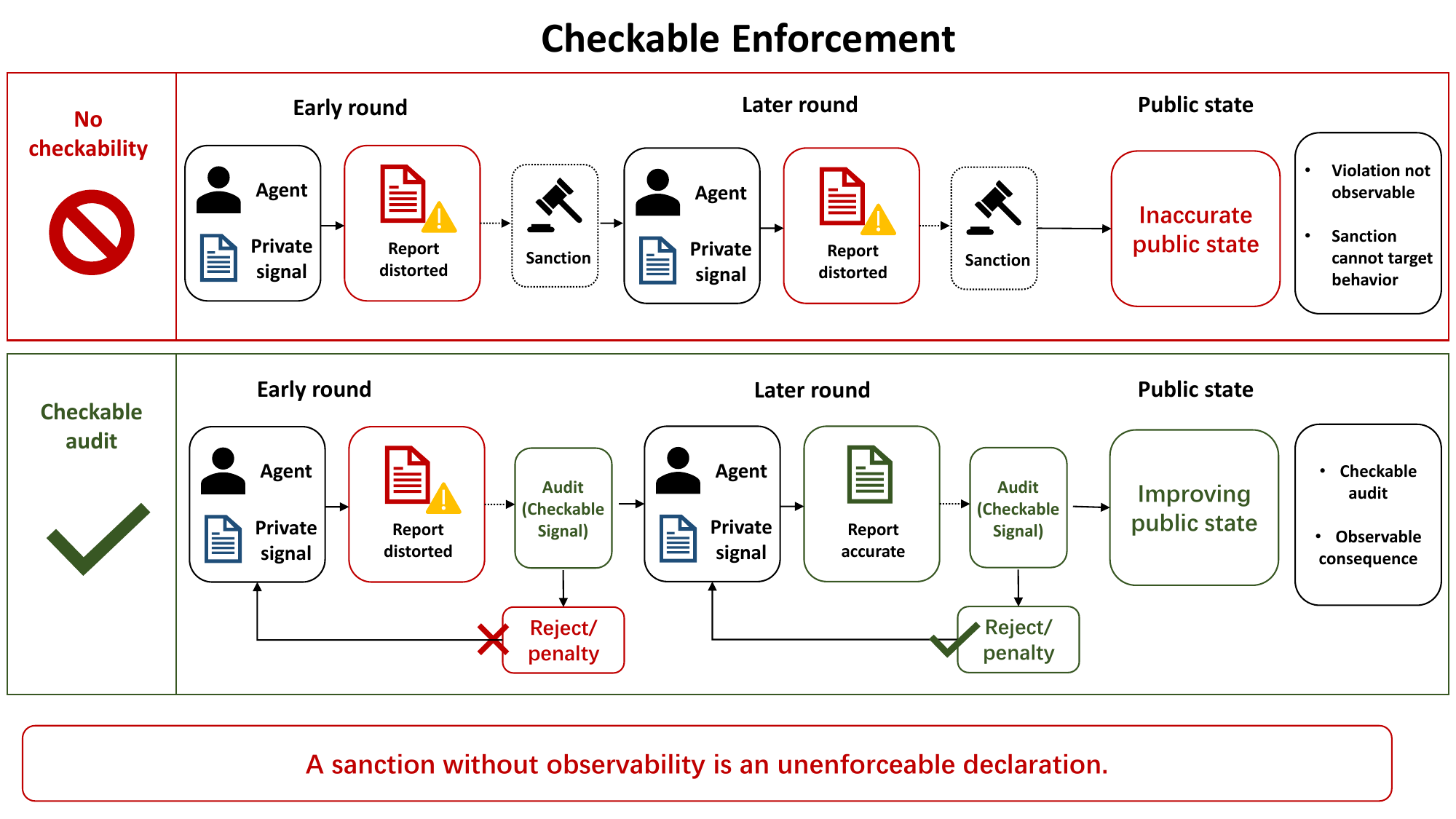}
  \caption{\textbf{Checkability makes enforcement behaviorally active.}
  The main panel depicts the simulated repeated-reporting mechanism: sanctions cannot condition on unobservable violations, whereas checkable audit can connect deviations to future consequences.
}
  \label{fig:accountability}
    \vspace{-0.3cm}
\end{figure}

\section{Accountability Through Time: Maintenance and Incentives}
\label{sec:accountability}

\paragraph{Group-psychology motivation.}
Collective knowledge must remain current and strategically credible.
Dynamic situation management requires teams to update shared representations as the task changes \cite{hoc2000}.
Accountability research also warns that merely demanding justification does not uniformly improve judgment: effects depend on what is observable, who evaluates, and how consequences are attached \cite{tetlock1989,tetlockboettger1989,schillemans2022}.
A sanction that cannot be connected to an observable violation is a declared norm, not an enforceable institution.

\paragraph{Artificial ecologies.}
Figure~\ref{fig:accountability} separates observability from consequence.
The dynamic-state environment gives agents chronological records in which a later observation supersedes an earlier one.
The institution maintains a current public state from a visible update relation; the broken regime degrades the visibility and reliability of that relation.
The repeated-reporting ecology introduces misaligned agents that may report a private capture option rather than their signal.
Audit probability determines whether reports are inspected, checkability determines whether inspection can identify a fabrication, and sanctions alter future incentives.
Private-signal capability supplies the intelligence comparison.
A separate hosted-policy pilot asks whether language-model reporters exhibit the same adaptive behavior.

\paragraph{Result and interpretation.}
Across Qwen 9B, Mistral 24B, and Qwen 235B, dynamic-state valid-minus-broken interactions are weak at 0.15, 0.15, and 0.30; only the Qwen 235B interval excludes zero (Fig.~\ref{fig:results-validity}a).
This heterogeneous boundary shows that public memory does not receive a generic benefit from added structure.

In the adaptive simulator, a 10\% audit at capability 0.55 raises accuracy from 0.265 to 0.535, reduces private capture from 0.676 to 0.398, and beats unaudited capability 0.65 by $+0.135$ [0.096, 0.175].
At zero checkability, every sanction has exactly 0.00 accuracy gain under common random numbers (Fig.~\ref{fig:results-validity}c).
The mechanism is not punishment by itself, but observable evidence that connects behavior to future consequence.

The hosted pilot marks the behavioral limit: across 2,400 Llama and Mistral decisions, both families reported honestly in every condition, so the predeclared behavioral gate stopped scaling (Table~\ref{tab:boundary-evidence}).
The selector's 0.022 regret on 16 disjoint simulator settings is a prospective within-ecology design result, not a universal policy.

\begin{figure}[htbp]
  \centering
  \includegraphics[width=\columnwidth]{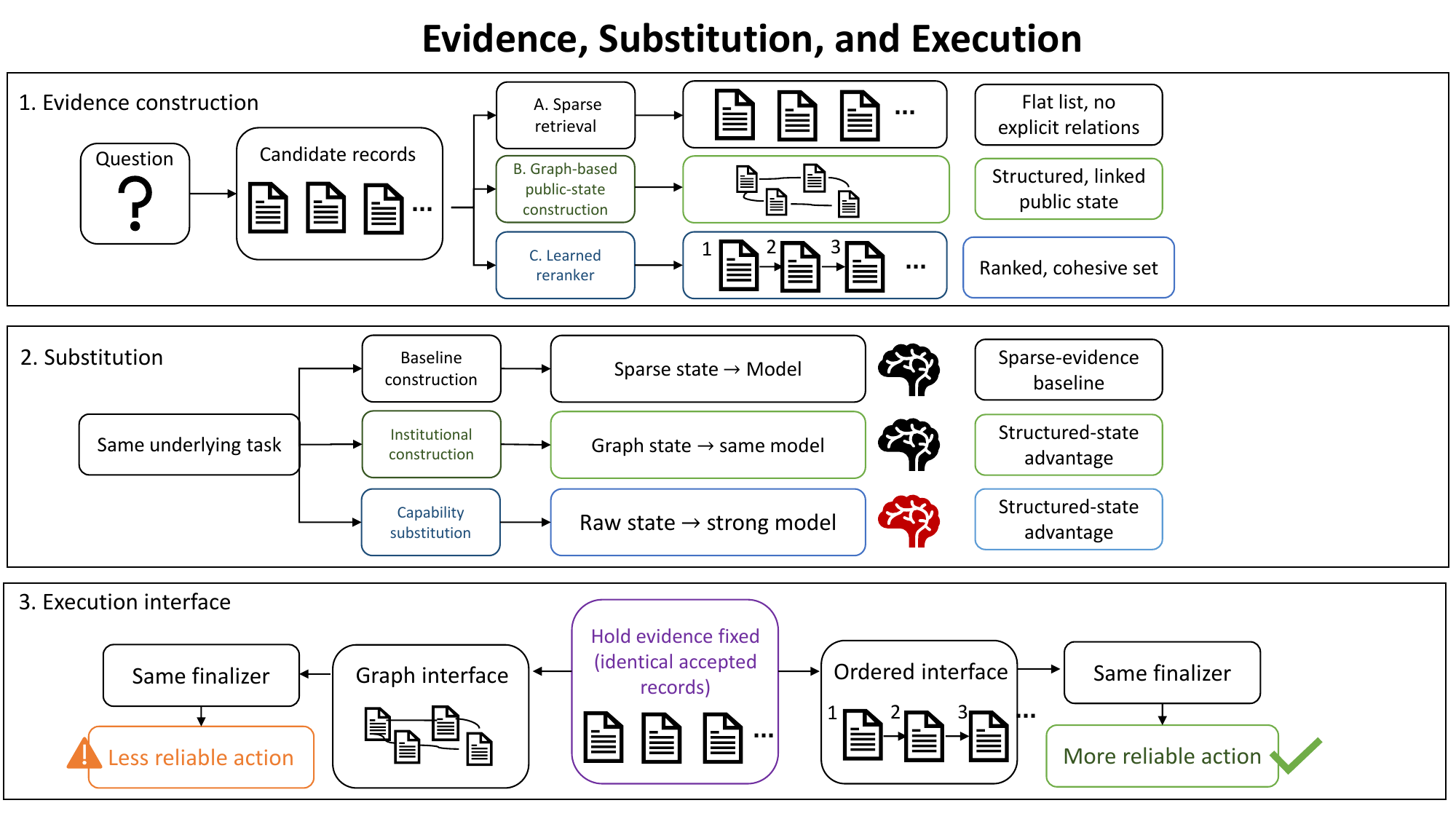}
  \caption{Construction, substitution, and execution are distinct. Learned components or stronger models may substitute for explicit institutional construction, while fixed-evidence comparisons isolate whether the resulting state is executable.}
  \label{fig:execution}
  \vspace{-0.3cm}
\end{figure}

\section{From Public Evidence to Public Action}
\label{sec:representation}

\paragraph{Distributed-cognition motivation.}
Collective reasoning does not end once relevant evidence has been collected.
Accounts of distributed cognition treat agents and external representations as parts of one functional system \cite{hutchins1995,michaelian2013}, and work on representation shows that two states can contain the same information while imposing very different operations on the decision maker \cite{zhangnorman1994,scaife1996}.
For an artificial collective, this creates three distinct questions.
Can an institution construct a better public evidence state?
Can a stronger model or learned component perform the same transformation without the institution?
And, even when the evidence is better, can the final agent reliably act on the way that evidence is represented?

\begin{figure*}[!t]
  \centering
  \includegraphics[width=1\textwidth]{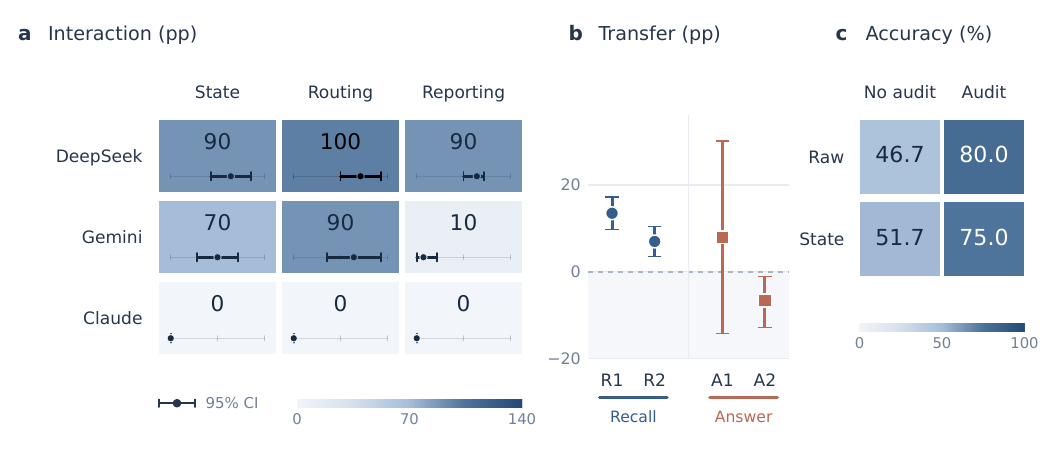}
    \vspace{-0.3cm}
    \caption{\textbf{Capability, construction, and the interface bound institutional advantage.}
    (a) Frontier-model validity interactions across public state, expertise routing, and strategic reporting (10 paired seeds per cell).
    (b) Support-recall contrasts: R1, entity graph minus BM25; R2, cross-encoder minus entity graph.
    Answer-accuracy contrasts: A1, graph context minus BM25 context; A2, public state minus ordered passages with evidence fixed.
    (c) Accuracy (\%) crossing audit with the finalizer interface; only the finalizer is hosted.
    Panels (a,b) report effects in pp with 95\% confidence intervals, clustered by question for A1 and source component for A2.
    Inset intervals in (a) share a within-cell 0--140 scale.}
  \label{fig:results-execution}
  \vspace{-0.3cm}
\end{figure*}

\paragraph{Can the institution construct a better public state?}
We first test whether explicit structure improves which evidence reaches the finalizer.
HotpotQA \cite{yang2018} compares an observable title-link graph with equal-size lexical selection at similar input-token counts.
MuSiQue \cite{trivedi2022} provides a harder construction test: on 300 balanced 2--4 hop examples, we compare an entity graph with BM25, dense retrieval, hybrid retrieval, and an all-passage learned reranker.
Gold support labels are used only after selection for retrieval diagnostics.
Graph routing transfers across three HotpotQA finalizers and improves MuSiQue support recall over equal-passage BM25 by $+0.135$ [0.098, 0.172].
Thus a hand-specified relational structure can recover evidence that sparse lexical retrieval misses.

\paragraph{When does capability substitute for explicit structure?}
A useful construction rule need not remain useful as capability improves.
On MuSiQue, the learned reranker exceeds the hand-specified graph by $+0.070$ [0.036, 0.104], showing that learned evidence selection can substitute for explicit graph construction.
We test the same principle at the model level with a separate frontier panel.
There, the raw model must reconstruct transparent dynamic-state, expertise-routing, and reporting transformations without receiving the corresponding institutional state.
DeepSeek retains large valid-minus-broken interactions, Gemini partially substitutes, and Claude produces zero incremental interaction in all three tested ecologies (Fig.~\ref{fig:results-execution}a).
These are not failed replications.
They identify a boundary on institutional value: when a capable reasoner can reliably perform the same transparent transformation from raw context, keeping that transformation as permanent institutional machinery becomes unnecessary.

\paragraph{Can the finalizer act on the constructed state?}
Better evidence construction is still not sufficient for better decisions.
Although the graph improves MuSiQue support recall over BM25, its hosted answer advantage over BM25 is imprecise.
We therefore hold the selected evidence fixed and change only how it is exposed to the finalizer.
The same passages are presented either through the graph-style public state or through an ordered or passage-pointer interface.
Under this fixed-evidence comparison, the ordered interface reverses the graph comparison (Fig.~\ref{fig:results-execution}b).
The evidence is present in both conditions; what changes is the operation required of the finalizer.
This identifies the action interface as a possible downstream bottleneck: a public state can contain better evidence without making the remaining decision easier to execute.

\paragraph{Where does the end-to-end gain come from?}
Frozen natural histories finally separate upstream evidence supply from downstream representation.
The same frozen visible-record pool is evaluated under a $2\times2$ design that crosses upstream audit with either a raw-record or public-state finalizer interface.
At the raw interface, a 0.25 checkable audit raises accuracy from 0.467 to 0.800, a paired supply effect of $+0.333$ [0.217, 0.450].
Holding the accepted records fixed, however, the downstream public-state transformation has no detectable simple effect either without audit ($+0.050$ [$-0.067$, 0.150]) or under audit ($-0.050$ [$-0.150$, 0.050]).
The full audit-plus-public-state pipeline gains $+0.283$ [0.150, 0.417] over no-audit raw execution (Fig.~\ref{fig:results-execution}c).
The detectable improvement is therefore associated with changing which evidence reaches the finalizer, rather than with reformatting the same accepted evidence alone.
Only the finalizer is hosted in this experiment; the adaptive reporters are simulated.

\section{When Do Institutions Beat Intelligence?}
\label{sec:synthesis}

The ecologies represent different failures, but their positive, null, and reversal results support three recurring empirical conditions.

\paragraph{The failure is social-structural rather than merely computational.}
Institutional value persists when stronger or call-matched reasoning receives the same missing, contaminated, correlated, stale, or strategically distorted state.
It disappears when a capable raw model or learned constructor performs the same transformation.
Institutional advantage is therefore relative to a capability frontier, not a permanent property of a workflow.

\paragraph{The institutional signal is valid and behaviorally active.}
A rule earns causal credit when performance follows the signal it claims to use and the gain disappears when the signal is broken or made uncheckable.
Grounded rather than syntactic validation, valid rather than broken credentials, and nonzero rather than zero checkability satisfy this requirement.
Labels, sanctions, and structure have no independent force.

\paragraph{The public state is nonredundant and executable.}
An institution can improve an intermediate representation without improving the final decision.
Fixed-evidence comparisons and formatting nulls show that construction and use must be evaluated separately.
The relevant object is the entire path from private observation to action, not the apparent quality of a public artifact in isolation.

These conditions are cross-ecology regularities, not a sequential certification algorithm or a universal theorem.
They imply a three-way design choice.
Buy \emph{intelligence} when capability directly performs the required transformation.
Build an \emph{institution} when the active structural failure survives scaling, the rule observes it reliably, and the resulting state supports action under explicit resource accounting.
Redesign the \emph{interface}, or choose neither intervention, when the rule is uninformative, redundant, or produces a state the finalizer cannot use.

\begin{table}[htbp]
\caption{\textbf{Boundary, null, and substitution evidence.}
Each probe limits where institutional advantage should be expected.}
\label{tab:boundary-evidence}
\centering
\footnotesize
\renewcommand{\arraystretch}{1.03}
\begin{tabularx}{\columnwidth}{@{}p{0.30\columnwidth}X@{}}
\toprule
Probe & Evidence and implication \\
\midrule

Dynamic-state validity
&
Interactions are 0.15, 0.15, and 0.30; only Qwen 235B excludes zero.
Public-state structure is not generically beneficial.
\\

Zero checkability
&
All four sanction levels yield exactly 0.00 gain.
Sanctions require observable violations.
\\

Hosted strategic pilot
&
0 misreports in 2,400 decisions.
The interface was exercised, but hosted strategic adaptation was not observed.
\\

Frontier substitution
&
Claude shows 0.00 incremental interaction in all three tested ecologies.
Transparent institutional rules can become capability-substitutable.
\\

Learned substitution
&
Reranker minus graph support recall is $+0.070$ [0.036, 0.104].
Hand-specified construction is not permanently privileged.
\\

Graph answer transfer
&
Graph minus BM25 is $+0.079$ [$-0.143$, 0.302].
Better support recovery need not yield a detectable answer gain.
\\

Fixed-evidence interface
&
Graph minus ordered accuracy is $-0.067$ [$-0.128$, $-0.010$].
Representation can reverse performance with evidence held fixed.
\\

Frozen-pool interface
&
Public-state effects are $+0.050$ [$-0.067$, 0.150] without audit and
$-0.050$ [$-0.150$, 0.050] with audit.
Reformatting the same accepted evidence has no detectable benefit.
\\

\bottomrule
\end{tabularx}
\end{table}

\section{Related Work}

\paragraph{Multi-agent institutions.}
Electronic institutions and normative multi-agent systems treat roles, protocols, admissible interactions, monitoring, and enforcement as system-level rules for coordinating autonomous agents
\cite{esteva2001formal,esteva2002islander,boella2006normative}.
Institutional analysis similarly studies how rules-in-use structure collective outcomes
\cite{ostrom1990,ostrom2009}.
Recent work extends this perspective to LLM collectives, showing that governance topology, market rules, and executable governance mechanisms can substantially change collective behavior
\cite{fei2026agents,chupilkin2026artificial, syrnikov2026institutional}.
These studies establish that institutional design can matter for artificial collectives.
Our focus is the comparative identification problem:
when does a structural intervention repair a binding collective failure relative to additional capability, and when can stronger or learned reasoning substitute for the same transformation?
We therefore pair institutional gains with matched capability baselines, mechanism-breaking controls, and explicit substitution tests.

\paragraph{Collective information processing.}
Hidden profiles, transactive memory, social influence, accountability, and external cognition explain how collective performance can diverge from individual competence
\cite{lu2012,stasser2000,becker2017,tetlock1989,zhangnorman1994}.
Rather than cite these traditions as analogy alone, we use each to motivate a separable experimental variable:
coverage, evidential dependence, checkability, state maintenance, or action affordance.

\paragraph{LLM multi-agent systems.}
Debate, role specialization, voting, retrieval, memory, and tool-mediated coordination are widely used in LLM-agent systems
\cite{chan2024,du2024,wu2024}.
Recent work also begins to separate organization, coordination, and collaboration protocol as independently configurable design dimensions
\cite{chen2026toward}.
Agent benchmarks assess broad competence or interactive behavior
\cite{liu2024,park2023}, while cooperative-AI work emphasizes coordination and common ground
\cite{dafoe2021}.
Our contribution is not another fixed workflow, organizational architecture, or aggregate benchmark.
Instead, every positive structural intervention is paired with a stronger or call-matched reasoner and a condition under which its proposed mechanism should fail or become redundant.

\section{Limitations and Scope}

Several causal ecologies are deliberately synthetic, and transcript-level corruption provides causal precision rather than open-domain fact checking.
The psychological literature supplies construct provenance, not evidence that language-model agents share human cognitive mechanisms.
Adaptive strategic behavior remains simulation evidence: the hosted audit pilot elicited no strategic variation, and the frozen-history study hosts the finalizer rather than the reporters.
Natural execution panels are smaller than the offline retrieval audit.
Results depend on provider implementations and model versions, and the selector is prospective only within one strategic ecology.
Call matching does not equal matching tokens, latency, monetary cost, or engineering burden; we use explicit resource accounting without claiming timeless provider-price rankings.
Finally, the diagnostic decomposition organizes evidence but does not constitute an identified structural theory.

\section{Conclusion}

A collective can contain enough intelligence and enough information yet fail to transform private evidence into public action.
Across ecologies of access, admission, maintenance, incentives, representation, and execution, institutions outperform additional intelligence when they repair a structural bottleneck that capability alone does not remove.
The nulls and reversals define the same phenomenon:
unobservable violations make sanctions inert, capable models and learned retrieval absorb transparent institutional functions, and unusable interfaces erase gains from better evidence.
The practical question is therefore not whether institutions or intelligence are generally superior.
It is where collective computation breaks, and which intervention changes that location most directly.

\FloatBarrier
\bibliographystyle{ACM-Reference-Format}
\bibliography{references}

@book{hutchins1995,
  author = {Hutchins, Edwin},
  title = {Cognition in the Wild},
  publisher = {MIT Press},
  year = {1995}
}

@article{michaelian2013,
  author = {Michaelian, Kourken and Sutton, John},
  title = {Distributed Cognition and Memory Research: History and Current Directions},
  journal = {Review of Philosophy and Psychology},
  year = {2013},
  volume = {4},
  pages = {1--24},
  doi = {10.1007/s13164-013-0131-x}
}

@article{stasser1985,
  author = {Stasser, Garold and Titus, William},
  title = {Pooling of Unshared Information in Group Decision Making: Biased Information Sampling During Discussion},
  journal = {Journal of Personality and Social Psychology},
  year = {1985},
  volume = {48},
  number = {6},
  pages = {1467--1478},
  doi = {10.1037/0022-3514.48.6.1467}
}

@article{stasser2000,
  author = {Stasser, Garold and Vaughan, Sandra I. and Stewart, Dennis D.},
  title = {Pooling Unshared Information: The Benefits of Knowing How Access to Information Is Distributed among Group Members},
  journal = {Organizational Behavior and Human Decision Processes},
  year = {2000},
  volume = {82},
  number = {1},
  pages = {102--116},
  doi = {10.1006/obhd.2000.2890}
}

@article{vanginkel2009,
  author = {van Ginkel, Wendy P. and van Knippenberg, Daan},
  title = {Knowledge about the Distribution of Information and Group Decision Making: When and Why Does It Work?},
  journal = {Organizational Behavior and Human Decision Processes},
  year = {2009},
  volume = {108},
  number = {2},
  pages = {218--229},
  doi = {10.1016/j.obhdp.2008.10.003}
}

@article{greitemeyer2003,
  author = {Greitemeyer, Tobias and Schulz-Hardt, Stefan},
  title = {Preference-Consistent Evaluation of Information in the Hidden Profile Paradigm},
  journal = {Journal of Personality and Social Psychology},
  year = {2003},
  volume = {84},
  number = {2},
  pages = {322--339},
  doi = {10.1037/0022-3514.84.2.322}
}

@article{becker2017,
  author = {Becker, Joshua and Brackbill, Devon and Centola, Damon},
  title = {Network Dynamics of Social Influence in the Wisdom of Crowds},
  journal = {Proceedings of the National Academy of Sciences},
  year = {2017},
  volume = {114},
  number = {26},
  pages = {E5070--E5076},
  doi = {10.1073/pnas.1615978114}
}

@article{freyrijt2021,
  author = {Frey, Vincenz and van de Rijt, Arnout},
  title = {Social Influence Undermines the Wisdom of the Crowd in Sequential Decision Making},
  journal = {Management Science},
  year = {2021},
  volume = {67},
  number = {7},
  pages = {4273--4286},
  doi = {10.1287/mnsc.2020.3713}
}

@article{almaatouq2022,
  author = {Almaatouq, Abdullah and Rahimian, M. Amin and Burton, Jason W. and Alhajri, Abdullah J.},
  title = {The Distribution of Initial Estimates Moderates the Effect of Social Influence on the Wisdom of the Crowd},
  journal = {Scientific Reports},
  year = {2022},
  volume = {12},
  doi = {10.1038/s41598-022-20551-7}
}

@inproceedings{hoc2000,
  author = {Hoc, Jean-Michel},
  title = {Cognitive Aspects of Dynamic Situation Management},
  booktitle = {Proceedings of the Human Factors and Ergonomics Society Annual Meeting},
  year = {2000},
  volume = {44},
  pages = {156--156},
  doi = {10.1177/154193120004400141}
}

@article{tetlock1989,
  author = {Tetlock, Philip E. and Skitka, Linda and Boettger, Richard},
  title = {Social and Cognitive Strategies for Coping with Accountability: Conformity, Complexity, and Bolstering},
  journal = {Journal of Personality and Social Psychology},
  year = {1989},
  volume = {57},
  number = {4},
  pages = {632--640},
  doi = {10.1037/0022-3514.57.4.632}
}

@article{schillemans2022,
  author = {Schillemans, Thomas},
  title = {Accountability and the Quality of Regulatory Judgment Processes},
  journal = {Public Performance \& Management Review},
  year = {2022},
  volume = {45},
  pages = {473--498},
  doi = {10.1080/15309576.2022.2040034}
}

@article{scaife1996,
  author = {Scaife, Mike and Rogers, Yvonne},
  title = {External Cognition: How Do Graphical Representations Work?},
  journal = {International Journal of Human-Computer Studies},
  year = {1996},
  volume = {45},
  number = {2},
  pages = {185--213},
  doi = {10.1006/ijhc.1996.0048}
}

@book{ostrom1990,
  author = {Ostrom, Elinor},
  title = {Governing the Commons: The Evolution of Institutions for Collective Action},
  publisher = {Cambridge University Press},
  year = {1990},
  doi = {10.1017/CBO9780511807763}
}

@article{ostrom2009,
  author = {Ostrom, Elinor},
  title = {A General Framework for Analyzing Sustainability of Social-Ecological Systems},
  journal = {Science},
  year = {2009},
  volume = {325},
  number = {5939},
  pages = {419--422},
  doi = {10.1126/science.1172133}
}

@inproceedings{chan2024,
  author = {Chan, Chi-Min and Chen, Weize and Su, Yusheng and Yu, Jianxuan and Xue, Wei and Zhang, Shanghang and Fu, Jie},
  title = {ChatEval: Towards Better LLM-Based Evaluators through Multi-Agent Debate},
  booktitle = {International Conference on Learning Representations},
  year = {2024}
}

@inproceedings{du2024,
  author = {Du, Yilun and Li, Shuang and Torralba, Antonio and Tenenbaum, Joshua B. and Mordatch, Igor},
  title = {Improving Factuality and Reasoning in Language Models through Multiagent Debate},
  booktitle = {Proceedings of the 41st International Conference on Machine Learning},
  year = {2024}
}

@article{dafoe2021,
  author = {Dafoe, Allan and others},
  title = {Cooperative AI: Machines Must Learn to Find Common Ground},
  journal = {Nature},
  year = {2021},
  volume = {593},
  pages = {33--36},
  doi = {10.1038/d41586-021-01170-0}
}

@inproceedings{liu2024,
  author = {Liu, Xiao and others},
  title = {AgentBench: Evaluating LLMs as Agents},
  booktitle = {International Conference on Learning Representations},
  year = {2024}
}

@inproceedings{park2023,
  author = {Park, Joon Sung and O'Brien, Joseph and Cai, Carrie Jun and Morris, Meredith Ringel and Liang, Percy and Bernstein, Michael S.},
  title = {Generative Agents: Interactive Simulacra of Human Behavior},
  booktitle = {Proceedings of the 36th Annual ACM Symposium on User Interface Software and Technology},
  year = {2023},
  doi = {10.1145/3586183.3606763}
}

@inproceedings{wu2024,
  author = {Wu, Qingyun and Bansal, Gagan and Zhang, Jieyu and Wu, Yiran and Li, Beibin and Zhu, Erkang and Jiang, Li and Zhang, Xiaoyun and Zhang, Shaokun and Liu, Jiale and Awadallah, Ahmed Hassan and White, Ryen W. and Burger, Doug and Wang, Chi},
  title = {AutoGen: Enabling Next-Gen LLM Applications via Multi-Agent Conversation},
  booktitle = {Conference on Language Modeling},
  year = {2024}
}

@inproceedings{yang2018,
  author = {Yang, Zhilin and Qi, Peng and Zhang, Saizheng and Bengio, Yoshua and Cohen, William W. and Salakhutdinov, Ruslan and Manning, Christopher D.},
  title = {HotpotQA: A Dataset for Diverse, Explainable Multi-Hop Question Answering},
  booktitle = {Proceedings of the 2018 Conference on Empirical Methods in Natural Language Processing},
  year = {2018},
  pages = {2369--2380},
  doi = {10.18653/v1/D18-1259}
}

@article{trivedi2022,
  author = {Trivedi, Harsh and Balasubramanian, Niranjan and Khot, Tushar and Sabharwal, Ashish},
  title = {MuSiQue: Multihop Questions via Single-Hop Question Composition},
  journal = {Transactions of the Association for Computational Linguistics},
  year = {2022},
  volume = {10},
  pages = {539--554},
  doi = {10.1162/tacl_a_00475}
}

@article{valckenaers2007,
  author = {Valckenaers, Paul and Sauter, John and Sierra, Carles and Rodr{\'i}guez-Aguilar, Juan A.},
  title = {Applications and Environments for Multi-Agent Systems},
  journal = {Autonomous Agents and Multi-Agent Systems},
  year = {2007},
  volume = {14},
  pages = {61--85},
  doi = {10.1007/s10458-006-9002-5}
}

@article{lu2012,
  author = {Lu, Li and Yuan, Y. Connie and McLeod, Poppy Lauretta},
  title = {Twenty-Five Years of Hidden Profiles in Group Decision Making: A Meta-Analysis},
  journal = {Personality and Social Psychology Review},
  year = {2012},
  volume = {16},
  number = {1},
  pages = {54--75},
  doi = {10.1177/1088868311417243}
}

@article{tetlockboettger1989,
  author = {Tetlock, Philip E. and Boettger, Richard},
  title = {Accountability: A Social Magnifier of the Dilution Effect},
  journal = {Journal of Personality and Social Psychology},
  year = {1989},
  volume = {57},
  number = {3},
  pages = {388--398},
  doi = {10.1037/0022-3514.57.3.388}
}

@article{zhangnorman1994,
  author = {Zhang, Jiajie and Norman, Donald A.},
  title = {Representations in Distributed Cognitive Tasks},
  journal = {Cognitive Science},
  year = {1994},
  volume = {18},
  number = {1},
  pages = {87--122},
  doi = {10.1207/s15516709cog1801_3}
}

@incollection{esteva2001formal,
  title={On the formal specification of electronic institutions},
  author={Esteva, Marc and Rodriguez-Aguilar, Juan-Antonio and Sierra, Carles and Garcia, Pere and Arcos, Josep L},
  booktitle={Agent Mediated Electronic Commerce: The European AgentLink Perspective},
  pages={126--147},
  year={2001},
  publisher={Springer}
}

@inproceedings{esteva2002islander,
  title={ISLANDER: an electronic institutions editor},
  author={Esteva, Marc and De La Cruz, David and Sierra, Carles},
  booktitle={Proceedings of the first international joint conference on Autonomous agents and multiagent systems: part 3},
  pages={1045--1052},
  year={2002}
}

@article{boella2006normative,
  title={Introduction to normative multiagent systems},
  author={Boella, Guido and Van Der Torre, Leendert and Verhagen, Harko},
  journal={Computational \& Mathematical Organization Theory},
  volume={12},
  number={2},
  pages={71--79},
  year={2006},
  publisher={Springer}
}

@article{fei2026agents,
  title={When Agents Evolve, Institutions Follow},
  author={Fei, Chao and Guo, Hongcheng and Xiao, Yanghua},
  journal={arXiv preprint arXiv:2604.27691},
  year={2026}
}

@article{chupilkin2026artificial,
  title={Artificial Institutions: How Institutional Design Shapes LLM Simulations},
  author={Chupilkin, Maxim},
  journal={arXiv preprint arXiv:2608.04020},
  year={2026}
}

@article{syrnikov2026institutional,
  title={Institutional AI: Governing LLM collusion in multi-agent cournot markets via public governance graphs},
  author={Syrnikov, Marcantonio Bracale and Pierucci, Federico and Galisai, Marcello and Prandi, Matteo and Bisconti, Piercosma and Giarrusso, Francesco and Sorokoletova, Olga and Suriani, Vincenzo and Nardi, Daniele},
  journal={arXiv preprint arXiv:2601.11369},
  year={2026}
}

@article{chen2026toward,
  title={Toward an Organizational Science of Multi-Agent LLM Systems: Decoupling Who, How, and Which Algorithm},
  author={Chen, Huan and Song, Xiang and Jin, Jian and Ren, Pan and Zhang, Liang-Jie},
  journal={arXiv preprint arXiv:2607.25446},
  year={2026}
}

\end{document}